\documentstyle[11pt]{article}
\newcommand{\beq}{\begin{equation}}
\newcommand{\eeq}{\end{equation}}

\newcommand{\beqs}{\begin{eqnarray}}
\newcommand{\eeqs}{\end{eqnarray}}
\newcommand{\beqsn}{\begin{eqnarray*}}
\newcommand{\eeqsn}{\end{eqnarray*}}

\newcommand{\nn}{\nonumber}

\begin{document}

\begin{center}
\vskip 2.5cm
{\LARGE \bf Proper incorporation of self-adjoint extension method to
Green's function formalism : one-dimensional $\delta^{'}$-function potential
case}

\vskip 1.0cm
{\Large  D.~K.~Park }
\\
{\large  Department of Physics,  KyungNam University, Masan, 631-701,
Korea}

\vskip 0.4cm
\end{center}

\centerline{\bf Abstract}

One-dimensional $\delta^{'}$-function potential is discussed in the framework
of Green's function formalism without invoking perturbation expansion.
It is shown that the energy-dependent Green's function for this case
is crucially dependent on the boundary conditions which are provided by
self-adjoint extension method. The most general Green's function which
contains four real self-adjoint extension parameters is constructed.
Also the relation between the bare coupling constant and self-adjoint
extension parameter is derived.
\vfill

\newpage
\setcounter{footnote}{1}

\newcommand{\tr}{\;{\rm tr}\;}
\section{Introduction}
Since Kronig-Penny model[1] has been successful for the description of
energy band in solid state physics, the point interaction problem has
been applied in the various branches of physics for a long time.
Recently the two-dimensional $\delta$-function potential has been of
interests in the context of the Aharonov-Bohm(AB) effect of spin-1/2
particles[2, 3] in which the delta function occurs as the mathematical
description of the Zeeman interaction of the spin with a magnetic flux
tube. In Ref.[4] two different approaches, renormalization and self-adjoint
extension methods[5, 6], are presented for this subject. More recently
same problem is re-examined in the framework of Green's function
formalism[7, 8]. In Ref.[8] present author showed how to incorporate
the self-adjoint extension method within the Green's function formalism
without invoking the perturbation expansion.

Unlike two- and three-dimensional cases, one-dimensional point interaction
provides a four-parameter family solution, characterized by the boundary
conditions at $x = 0$ :
\beqs
    \varphi(\epsilon)&=& \omega a \varphi(-\epsilon) + \omega b
    \varphi^{'}(- \epsilon),  \\  \nn
    \varphi^{'}(\epsilon)&=& \omega c \varphi(-\epsilon) + \omega d
    \varphi^{'}(-\epsilon),
\eeqs
where $\epsilon$ is infinitesimal positive parameter and $\omega \in {\bf C}$;
$a, b, c, d \in {\bf R}$, satisfying $\mid\omega\mid = 1$ and
$ ad - bc = 1 $[6, 9].
Recently path-integral for the one-dimensional $\delta^{'}$-function
potential is calculated by incorporating Neumann boundary conditions
within the usual perturbation theory of one-dimensional Dirac particle
in order for the coupling constant to be infinitely repulsive[10].

In this paper we will discuss the one-dimensional $\delta^{'}$-function
potential in the framework of Green's function formalism without using
a perturbation expansion like Ref.[8].
It will be shown that
the energy-dependent Green's function is crucially dependent on the
boundary conditions which are provided by self-adjoint extension method
in the present formalism. Choosing the boundary condition
\beqs
\varphi^{'}(\epsilon) = \varphi^{'}(- \epsilon)&=& \varphi^{'}(0),  \\ \nn
\varphi(\epsilon) - \varphi(-\epsilon)         &=& \beta \varphi^{'}(0),
\eeqs
which is easily obtained from Eq.(1) by requring $c = 0, \omega = a = d = 1,$
and $ b = \beta $, one can derive a similar result with that of Ref.[10].

However, the advantage of this formalism presented here is that it is free
to choose boundary conditions. This means that one can get more general
Green's function by choosing more general boundary conditions. If one
chooses the most general boundary conditions (1) of one-dimensional
point interaction, the most general
Green's function, in which four real self-adjoint extension parameters
are contained, can be derived.
It is worthwhile to note that this formalism does not use
the complicated perturbation expansion. Therefore, calculation is
very simple and clear.

This paper is organized as follows. In Sec.2 we will discuss why the
derivation of Green's function for one-dimensional $\delta^{'}$-function
potential case is difficult. In this section we will show that the
representation of $\delta^{'}$-function as two usual $\delta$-function
with infinitesimal distance does generate the physically irrelevant Green's
function.
In Sec.3 we will derive energy-dependent Green's function by incorporating
the self-adjoint method in the Green's function formalism. In this section
we will show that the most general Green's function for one-dimensional
point interaction can be constructed by using only $\delta^{'}$-function
potential. In Sec.4 a brief conclusion is given.

\section{Calculational Difficulty of Green's function for one-dimensional
$\delta^{'}$-function potential}

In this section we will calculate the energy-dependent Brownian motion
Green's function $\hat{G}[x, y; E]$ for one-dimensional $\delta^{'}$-function
potential by using a same method presented in Ref.[11], in which the
one-dimensional $\delta$-function potential case is calculated.
For this purpose consider a one-dimensional system whose Hamiltonian is
\beq
H = H_0 + v \delta^{'}(x),
\eeq
where $v$ is bare coupling constant.
Although $H_0$ can involve an arbitary potential, in this paper we will
only consider the free particle case for simplicity:
\beq
H_0 = \frac{p^2}{2}.
\eeq
It is well-known that the time-dependent Brownian motion propagator
for the Hamiltonian (3) obeys integral equation[12, 13]
\beq
G[x, y; t] = G_0[x, y; t] - v \int_{0}^{t} ds \int dz G_0[x, z; t-s]
             \delta^{'}(z) G[z, y; s].
\eeq
In order to follow the same method used in Ref.[11] for $\delta$-function
potential case, we regard $\delta^{'}$-function as
\beq
\delta^{'}(z) = \lim_{\epsilon \rightarrow 0^+}
                \frac{\delta(z + \epsilon) - \delta(z - \epsilon)}
                     {2 \epsilon}.
\eeq
After inserting Eq.(6) into (5), one can derive easily
\beqs
\hat{G}[x, y; E]&=& \hat{G}_0[x, y; E]   \\   \nn
                &+& \frac{v}{2 \epsilon}
                    \left[ \hat{G}_0[x, \epsilon; E] \hat{G}[\epsilon, y; E] -
                           \hat{G}_0[x, -\epsilon; E] \hat{G}[-\epsilon, y; E]
                    \right]
\eeqs
by taking Laplace transform
\beq
\hat{f}(E) \equiv \int_{0}^{\infty} dt e^{-Et} f(t)
\eeq
of both sides of Eq.(5).
At this stage the limit of $\epsilon$ is omitted for brevity. We will take
this limit after calculation. After inserting $ x = \pm \epsilon $ in Eq.(7),
one can obtain with much ease
\beqs
\hat{G}[\epsilon, y; E]&=&
        \frac{ \left( 1 + \frac{v}{2 \epsilon \sqrt{2E}} \right)
               \hat{G}_0[\epsilon, y; E]
               - \frac{v e^{-\sqrt{2E}2 \epsilon}}{2 \epsilon \sqrt{2 E}}
                 \hat{G}_0[-\epsilon, y; E]
             }
             { 1 - \frac{v^2}{8 E \epsilon^2}
                   \left( 1 - e^{- \sqrt{2E} 4 \epsilon} \right)
             }     \\     \nn
\hat{G}[- \epsilon, y; E]&=&
          \frac{\frac{v e^{-\sqrt{2E}2 \epsilon}}{2 \epsilon \sqrt{2 E}}
                \hat{G}_0[\epsilon, y; E]
                + \left( 1 - \frac{v}{2 \epsilon \sqrt{2E}} \right)
                \hat{G}_0[-\epsilon, y; E]
               }
               {1 - \frac{v^2}{8 E \epsilon^2}
                   \left( 1 - e^{- \sqrt{2E} 4 \epsilon} \right)
               }.
\eeqs
When deriving Eq.(9) we used the explicit result of energy-dependent
Green's function for one-dimensional free particle
\beq
\hat{G}_0[x, y; E] = \frac{1}{\sqrt{2E}}
                     e^{-\sqrt{2E} \mid x - y \mid}.
\eeq
By inserting Eq.(9) into Eq.(7) $\hat{G}[x, y; E]$ becomes
\beqs
\hat{G}[x, y; E] &=& \hat{G}_0[x, y; E]
                 + \frac{v}{4 \pi \epsilon}
                     \frac{1}
                          {1 - \frac{v^2}{8 E \epsilon^2}
                   \left( 1 - e^{- \sqrt{2E} 4 \epsilon} \right)} \\ \nn
                &\times&
                 \Bigg[
                       \left( 1 + \frac{v}{2 \epsilon \sqrt{2E}} \right)
                       e^{- \sqrt{2E} ( \mid x - \epsilon \mid
                                       + \mid y - \epsilon \mid
                                      )
                         }
                       - \frac{v e^{ - \sqrt{2E} 2 \epsilon}}
                               {2 \epsilon \sqrt{2E}}
                       e^{- \sqrt{2E} ( \mid x - \epsilon \mid
                                       + \mid y + \epsilon \mid
                                      )
                         }  \\ \nn
                       &-& \frac{v e^{ - \sqrt{2E} 2 \epsilon}}
                               {2 \epsilon \sqrt{2E}}
                       e^{- \sqrt{2E} ( \mid x + \epsilon \mid
                                       + \mid y - \epsilon \mid
                                      )
                         }
                       - \left( 1 - \frac{v}{2 \epsilon \sqrt{2E}} \right)
                       e^{- \sqrt{2E} ( \mid x + \epsilon \mid
                                       + \mid y + \epsilon \mid
                                      )
                         }
                 \Bigg].
\eeqs
Now let us assume that the coupling constant $v$ is $\epsilon$-independent
and finite.
If one calculates $\hat{G}[x, y; E]$ by taking $ \epsilon \rightarrow 0^+$
limit in Eq.(11) at the following four regions
\beqs
x&>&\epsilon    \hspace{.5in} y\hspace{0.05in}>\hspace{0.05in}  \epsilon, \\
\nn
x&>& \epsilon   \hspace{.5in} y\hspace{0.05in} <\hspace{0.05in} -\epsilon,\\
\nn
x&<& -\epsilon  \hspace{.5in}  y\hspace{0.05in} >\hspace{0.05in}  \epsilon, \\
 \nn
x&<& -\epsilon  \hspace{.5in}  y\hspace{0.05in} <\hspace{0.05in} -\epsilon,
\eeqs
one can show that the infinity terms( ${\cal O}(\epsilon^{-1}) $ ) cancel,
leaving the same finite results at these four regions
\beq
\hat{G}[x, y; E] = \hat{G}_0[x, y; E] - \frac{1}{\sqrt{2E}}
                    e^{-\sqrt{2E} ( \mid x \mid + \mid y \mid )}.
\eeq
At this stage the regions $ -\epsilon < x < \epsilon $ and
$ -\epsilon < y < \epsilon $ are not considered since both shrink
infinitesimally if $ \epsilon \rightarrow 0^+$ limit is taken.

The result (13) is physically irrelevant since it is independent of
coupling constant $v$. There is another reason which makes Eq.(13)
physically unacceptable. From Ref.[11] the energy-dependent Brownian
motion Green's
function for one-dimensional $\delta$-function potential case is
\beq
\hat{G}_{\delta}[x, y; E] = \hat{G}_0[x, y; E] -
                            \frac{c}{\sqrt{2E}(\sqrt{2E} + c)}
                            e^{-\sqrt{2E} ( \mid x \mid + \mid y \mid)}
\eeq
where $c$ is coupling constant of $\delta$-function potential.
Then one can easily show
\beq
\hat{G}[x, y; E] = \lim_{c \rightarrow \infty}
                   \hat{G}_{\delta}[x, y; E]
\eeq
which results in the physically irrelevant deduction
\beqsn
v \delta^{'}(x) \stackrel{?}{=} \lim_{c \rightarrow \infty} c \delta(x) \nn
\eeqsn
at quantum level. Thus the description of $\delta^{'}$-function as two
usual $\delta$-function with infinitesimal distance does not make sense
physically if $v$ is $\epsilon$-independent and finite. Maybe some relations
between $v$ and $\epsilon$ can give  physically relevant solutions.
Upon my knowledge it is not clear how to derive the relations systematically.

In next section we will present the correct procedure for the calculation
of energy-dependent Green's function when the potential is one-dimensional
$\delta^{'}$-function.

\section{Green's function approach to one-dimensional $\delta^{'}$-function
potential}

In this section we will show how to incorporate the self-adjoint extension
method into Green's function formalism by using one-dimensional
$\delta^{'}$-function potential without invoking a perturbation expansion.
The two- and three-dimensional cases are already discussed in Ref.[8].

Now let us start with Eq.(5). After performing integration with respect to $z$
in Eq.(5), one can show easily
\beqs
\hat{G}[x, y; E]&=& \hat{G}_0[x, y; E]   \\    \nn
                &+& v \left(
                            \frac{\partial \hat{G}_0[x, z; E]}{\partial z}
                      \right)_{z = 0} \hat{G}[0, y; E]  \\    \nn
                &+& v \hat{G}_0[x, 0; E]
                      \left(
                            \frac{\partial \hat{G}[z, y; E]}{\partial z}
                      \right)_{z=0}.
\eeqs
Eq.(16) is purely formal. This is easily deduced from fact that
$\hat{G}[0, y; E]$ is not well-defined because of the factor $\mid x \mid$
which is contained in $(\partial \hat{G}_0[x, z; E] / \partial z)_{z = 0}$.
Therefore, at this stage one has to conjecture the modification of Eq.(16).
Our conjecture for the modification of Eq.(16) is simply to extract the
problematic zero point at $\hat{G}[x, y; E]$ as follows:
\beqs
\hat{G}[x, y; E]&=& \hat{G}_0[x, y; E]   \\    \nn
                &+& v \left(
                            \frac{\partial \hat{G}_0[x, z; E]}{\partial z}
                      \right)_{z = 0} \hat{G}[\epsilon, y; E] \hspace{.3in} for
\hspace{.1in} x > 0 \\ \nn
                &+& v \hat{G}_0[x, 0; E]
                      \left(
                            \frac{\partial \hat{G}[z, y; E]}{\partial z}
                      \right)_{z= \epsilon}  \\ \nn
                                             \\ \nn
\hat{G}[x, y; E]&=& \hat{G}_0[x, y; E]   \\    \nn
                &+& v \left(
                            \frac{\partial \hat{G}_0[x, z; E]}{\partial z}
                      \right)_{z = 0} \hat{G}[-\epsilon, y; E] \hspace{.3in}
for\hspace{.1in} x<0  \\ \nn
                &+& v \hat{G}_0[x, 0; E]
                      \left(
                            \frac{\partial \hat{G}[z, y; E]}{\partial z}
                      \right)_{z= -\epsilon}
\eeqs
which might be natural modification of Eq.(16).

In Eq.(17) the infinitesimal positive parameter $\epsilon$ is introduced again.
By inserting $ x = \pm \epsilon $ in the first and second equations of Eq.(17)
respectively, one can derive
\beqs
\left(
      \frac{\partial \hat{G}[z, y; E]}{\partial z}
\right)_{z= \epsilon}&=& \frac{\sqrt{2E}}{v}
                        \left[ (1 - v) \hat{G}[\epsilon, y; E] -
                               \hat{G}_0[0, y; E]
                        \right]     \\  \nn
\left(
      \frac{\partial \hat{G}[z, y; E]}{\partial z}
\right)_{z= - \epsilon}&=& \frac{\sqrt{2E}}{v}
                           \left[ (1 + v) \hat{G}[-\epsilon, y; E] -
                                  \hat{G}_0[0, y; E]
                           \right].
\eeqs
By inserting Eq.(18) into Eq.(17) $\hat{G}[x, y; E]$ becomes
\beqs
\hat{G}[x, y; E]
&=& \hat{G}_0 [x, y; E]  \\  \nn
&-& \frac{1}{\sqrt{2E}}
                                          e^{-\sqrt{2E}(\mid x \mid +
                                             \mid y \mid)}
                                         + e^{-\sqrt{2E} \mid x \mid}
                                           \hat{G}[\epsilon, y; E]
                               \hspace{.3in} for \hspace{.1in} x > 0 \\    \nn
\hat{G}[x, y; E] &=& \hat{G}_0 [x, y; E] \\  \nn
&-& \frac{1}{\sqrt{2E}}
                                          e^{-\sqrt{2E}(\mid x \mid +
                                             \mid y \mid)}
                                         + e^{-\sqrt{2E} \mid x \mid}
                                           \hat{G}[-\epsilon, y; E]
                          \hspace{.3in} for \hspace{.1in} x < 0.
\eeqs
Note that in Eq.(19) the $v$-dependence of $\hat{G}[x, y; E]$ is hidden
in $\hat{G}[\pm \epsilon, y; E]$.

Now it is time to incorporate the self-adjoint extension method into
Green's function formalism. Firstly let us consider the simple boundary
conditions given in Eq.(2). By applying these two boundary conditions
to $\hat{G}[x, y; E]$, one can show that the boundary conditions generate
two independent equations
\beqs
\hat{G}[\epsilon, y; E] + \hat{G}[-\epsilon, y; E]&=&
\frac{2}{\sqrt{2E}} e^{-\sqrt{2E} \mid y \mid}       \\  \nn
\hat{G}[\epsilon, y; E] - \hat{G}[-\epsilon, y; E]&=&
\beta \left[ ( \epsilon(y) + 1 ) e^{- \sqrt{2E} \mid y \mid}
             - \sqrt{2E} \hat{G}[\epsilon, y; E]
      \right]
\eeqs
where $\epsilon(y)$ is usual alternating function. Therefore,
by solving Eq.(20) the solutions
\beqs
\hat{G}[\epsilon, y; E] = \frac{1}{\sqrt{2E}} e^{-\sqrt{2E} \mid y \mid}
                          \left( 1 +
                                 \frac{\sqrt{2E}}{\sqrt{2E} + \frac{2}{\beta}}
                                 \epsilon(y)
                          \right),  \\   \nn
\hat{G}[-\epsilon, y; E] = \frac{1}{\sqrt{2E}} e^{-\sqrt{2E} \mid y \mid}
                           \left( 1 -
                                  \frac{\sqrt{2E}}{\sqrt{2E} + \frac{2}{\beta}}
                                 \epsilon(y)
                          \right)
\eeqs
are easily obtained. By combining Eqs.(19) and (21) we get a final result
\beq
\hat{G}[x, y; E] = \hat{G}_0[x, y; E] +
                   \frac{\epsilon(x) \epsilon(y)}
                        {\sqrt{2E} + \frac{2}{\beta}}
                   e^{-\sqrt{2E} ( \mid x \mid + \mid y \mid )}.
\eeq
Also the relation between the bare coupling constant $v$ and self-adjoint
extension parameter $\beta$ is obtained by inserting Eq.(21) into Eq.(18)
and using the continuity of $\partial \hat{G}[z, y; E] / \partial z$ at
$ z = 0 $. Unlike two- and three-dimensional cases the relation is
dependent on the space:
\beqs
\frac{1}{v}&=& \left( 1 + \sqrt{\frac{2}{E}} \frac{1}{\beta} \right)
             \hspace{.3in} for \hspace{.1in} y > 0,   \\  \nn
\frac{1}{v}&=& - \left( 1 + \sqrt{\frac{2}{E}} \frac{1}{\beta} \right)
              \hspace{.3in} for \hspace{.1in} y < 0.
\eeqs
After taking inverse Laplace transform of Eq.(22) and using analytic
continuation in time, one can obtain Feynman propagator(or Kernel)
$K[x, y; t]$:
\beqs
K[x, y; t]&=& \frac{1}{\sqrt{2 \pi i t}}
             exp \left( \frac{i}{2t} \mid x - y \mid^{2} \right)  \\  \nn
          &+& \frac{1}{\sqrt{2 \pi i t}}
             exp \left( \frac{i}{2t} ( \mid x \mid + \mid y \mid )^2 \right)
             \epsilon(x) \epsilon(y)   \\   \nn
          &-& \frac{1}{\beta}
             exp \left( \frac{2}{\beta} (\mid x \mid + \mid y \mid )
                        + \frac{2it}{\beta^{2}}
                 \right) \\    \nn
          &\times& erfc
                   \left[ \frac{1}{\sqrt{2it}}[ \mid x \mid + \mid y \mid
                                                + \frac{2it}{\beta}
                                              ]
                   \right] \epsilon(x) \epsilon(y),
\eeqs
where $erfc(z)$ is usual error function.
Eq.(24) coincides with Eq.(14) of Ref.[10] if one changes $\beta$ of Ref.[10]
into $ -\beta / 2$. Therefore we derived a similar result with that of
Ref.[10] without invoking perturbation expansion. Furthermore, in this
formalism one can derive more general Green's function(or propagator) by
using more general boundary conditions. Therefore, let us use the most
general boundary condition (1) of one dimensional point interaction.
Like same way as before these two boundary conditions provide two
independent equations
\beqs
(\frac{d}{b} + \sqrt{2E}) \hat{G}[\epsilon, y; E] -
\frac{\omega}{b} \hat{G}[-\epsilon, y; E]&=& (\epsilon(y) + 1)
                                             e^{-\sqrt{2E} \mid y \mid} \\ \nn
\frac{\omega^{\star}}{b} \hat{G}[\epsilon, y; E] -
(\sqrt{2E} + \frac{a}{b}) \hat{G}[-\epsilon, y; E]&=& (\epsilon(y) - 1)
                                                     e^{-\sqrt{2E} \mid y \mid}
\eeqs
where $\omega^{\star}$ is complex conjugate of $\omega$.
By inserting the solutions of Eq.(25)
\beqs
& &\hat{G}[\epsilon, y; E]   \\   \nn
&=&
                -\frac{e^{-\sqrt{2E} \mid y \mid}}
                      {\frac{c}{b} + \sqrt{2E} \frac{a+d}{b} + 2E}
                \Bigg[ \epsilon(y) \left( \frac{\omega}{b} - \sqrt{2E}
                                          - \frac{a}{b}
                                   \right) -
                                   \left( \frac{\omega}{b} + \sqrt{2E}
                                          + \frac{a}{b}
                                   \right)
                \Bigg]  \\  \nn
& &\hat{G}[-\epsilon, y; E]  \\  \nn
&=&
                 -\frac{e^{-\sqrt{2E} \mid y \mid}}
                      {\frac{c}{b} + \sqrt{2E} \frac{a+d}{b} + 2E}
                 \Bigg[ \epsilon(y) \left( \frac{d}{b} + \sqrt{2E}
                                           -\frac{\omega^{\star}}{b}
                                    \right) -
                                    \left( \frac{d}{b} + \sqrt{2E}
                                          + \frac{\omega^{\star}}{b}
                                    \right)
                 \Bigg]
\eeqs
to Eq.(19) it is straightforward to derive the energy-dependent Green's
 function corresponding to the most general boundary conditions
\beqs
\hat{G}[x, y; E]&=& \hat{G}_0[x, y; E] +
                    \frac{\sqrt{2E} b}{D(E)} e^{-\sqrt{2E} (\mid x \mid
                                                + \mid y \mid )
                                               } \epsilon(x) \epsilon(y) \\ \nn
                &-& \frac{e^{-\sqrt{2E} ( \mid x \mid + \mid y \mid )}}
                         {D(E)}
                            \Bigg[ \frac{c}{\sqrt{2E}} +
                                   \frac{1}{2}(a + d - \omega - \omega^{\star})
\\ \nn
                                &+& \frac{1}{2}(d - a + \omega^{\star} -
\omega)
                                   \epsilon(x)
                                 + \frac{1}{2}(d - a + \omega - \omega^{\star})
                                   \epsilon(y)  \\  \nn
                                &-& \frac{1}{2}(a + d - \omega -
\omega^{\star})
                                 \epsilon(x) \epsilon(y)
                            \Bigg]
\eeqs
where
\beq
D(E) = c + (a + d) \sqrt{2E} + 2Eb.
\eeq
Note that Eq.(27) coincides with Eq.(22) at $c = 0$, $\omega = a = d = 1$,
and $b = \beta$. The energy-dependent Green's function for the one-dimensional
point interaction is calculated in Ref.[14]. The result (27) is exactly
same with that of Ref.[14] although the authors of Ref.[14] clamed that
their result is a consequence of appropriate mixture of one-dimensional
$\delta-$ and $\delta^{'}-$potentials. In this paper same result can be
derived by using only one-dimensional $\delta^{'}-$function potential. Of
course by following the procedure presented in Ref.[14] one can also obtain
the time-dependent Brownian motion propagator and
Feynman Kernel straightforwardly.
Also one can derive the relation between bare coupling constant and
self-adjoint parameters as before.

\section{Conclusion}
The one-dimensional $\delta^{'}-$function potential is analyzed in the
framework of Green's function formalism. It is shown that the energy-dependent
Brownian motion Green's function for one-dimensional
$\delta^{'}-$function potential is
crucially dependent on the boundary conditions. By choosing the most
general boundary condition of one-dimensional point interaction the most
general Green's function which contains four real self-adjoint extension
parameters is constructed. Grosch's result which is obtained recently
by using the perturbation expansion of one-dimensional Dirac particle is
special case of ours. Also the relation between the bare coupling constant
and self-adjoint extension parameters is derived. Unlike two- and
three-dimensional cases the relation is dependent on the space.


\begin{thebibliography}{9}

\bibitem{1} R. de L. Kronig and W. G. Penny, Proc. Roy. Soc. {\bf 130A},
499(1931).
\bibitem{2} Ph. de Sousa Gerbert, Phys. Rev. {\bf D40}, 1346(1989).
\bibitem{3} C.R.Hagen, Phys. Rev. Lett. {\bf 64}, 503(1990); Int. J. Mod. Phys.
{\bf A6}, 3119(1991).
\bibitem{4} R. Jackiw, in M. A. B\'{e}g Memorial Volume, edited by A. Ali and
P. Hoodbhoy(World Scientific, Singapore, 1991).
\bibitem{5} M. Reed and B. Simon, Methods of Modern Mathematical
Physics(Academic, New York, 1975).
\bibitem{6} S. Albeverio, F. Gesztesy, R. Hoegh-Krohn, and H. Holden, Solvable
Models in Quantum Mechanics(Springer, Berlin, 1988).
\bibitem{7} C. Grosche, Ann. Physik {\bf 3}, 283(1994).
\bibitem{8} D. K. Park, J. Math. Phys. {\bf 36}, 5453(1995).
\bibitem{9} P. R. Chernoff and R. J. Hughes, J. Funct. Anal. {\bf 111},
97(1993).
\bibitem{10} C. Grosche, J. Math. Phys. {\bf 28}, L99(1995).
\bibitem{11} B. Gaveau and L. S. Schulman, J. Phys. {\bf A19}, 1833(1986).
\bibitem{12} R. P. Feynman and A. R. Hibbs, Quantum Mechanics and Path
Integrals(McGraw-Hill, New York, 1965).
\bibitem{13} L. S. Schulman, Techniques and Applications of Path
Integrals(Wiley, New York, 1981).
\bibitem{14} S. Albeverio, Z. Brzezniak, and L. Dabrowski, J. Math. Phys. {\bf
A27}, 4933(1994).
\end{thebibliography}
\end{document}